\documentclass[aps,prb,longbibliography,twocolumn]{revtex4-2}

\usepackage{graphicx}
\usepackage{dcolumn}
\usepackage{textcomp}
\usepackage[T1]{fontenc}
\usepackage{times}
\usepackage[colorlinks = true, linkcolor = blue, urlcolor  = blue, citecolor = blue, anchorcolor = blue]{hyperref}
\usepackage[english]{babel}
\usepackage{multirow}
\usepackage{dcolumn}

\usepackage{titlesec}
\titlespacing\section{0pt}{18pt plus 0pt minus 0pt}{6pt plus 0pt minus 0pt}
\titlespacing\subsection{0pt}{12pt plus 0pt minus 0pt}{6pt plus 0pt minus 0pt}

\begin{document}

\title {Anisotropic Pseudospin Tunneling in Two-Dimensional Black Phosphorus Junctions}
\author{Young Woo Choi and Hyoung Joon Choi}
\email{h.j.choi@yonsei.ac.kr}
\address{Department of Physics, Yonsei University, Seoul 03722, Korea}

\begin{abstract}
  We investigate the role of pseudospin structure of few-layer black phosphorus (BP) in interband tunneling properties in lateral BP junctions.
  We find that interband tunneling is critically dependent on junction directions because of the anisotropic pseudospin structure of BP.
  When the armchair direction of BP is normal to the interface,
  pseudospins of incident and transmitted carriers are nearly aligned so that interband tunneling is highly effective,
  analogous to the Klein tunneling in graphene.
  However, when the zigzag direction is normal to the interface,
  interband tunneling is suppressed by misaligned pseudospins.
  We also study junctions of band-gap inverted BP where the electronic structure is characterized by two Dirac cones.
  In this case, intervalley tunneling is prohibited either by momentum conservation or by pseudospin mismatch
  while intravalley tunneling is Klein-like irrespective of the junction direction.
  These results provide a foundation for developing high-performance devices from BP and other pseudospin materials. \\

  \noindent Keywords: pseudospin transport, Klein tunneling, black phosphorus, two-dimensional materials
\end{abstract}

\maketitle

\section{Introduction}
Two-dimensional black phosphorus (BP),
a tunable gapped semi-Dirac material,
has attracted enormous attention
because of its unique electronic structure and outstanding properties
for high-performance electronic devices
\cite{Zhang:2014,Jia:2014,Dresselhaus:2015,Wu:2019}.
The band gap of few-layer black phosphorus sensitively depends on the number of
layers, ranging from 0.3 eV in its bulk form to 1.8 eV in
monolayer \cite{Ji:2014,Wang:2017,Yan:2017}. This widely tunable band gap serves as the
foundation for numerous promising device applications.

Recently, tunneling field effect transistors (TFET) based on
bulk-monolayer BP junctions have shown superb device performance in terms of
both switching speed and tunneling current \cite{Cho:2020a,Cho:2020b}.
In these works, high tunneling current was partly attributed
to the absence of interfacial problems.
However, in addition to such extrinsic factors,
it is highly likely that BP's intrinsic electronic structure can play an
important role in the interband tunneling properties of BP junctions.

In fact,
few-layer black phosphorus has a unique electronic structure
that can be characterized as a tunable gapped semi-Dirac spectrum
\cite{Pickett:2009,Vanderbilt:2015}.
BP has highly anisotropic energy spectrum that is
parabolic in the zigzag direction, but hyperbolic in the armchair direction
\cite{Ji:2014,Xia:2015}.
In addition, angle-resolved photoemission spectroscopy (ARPES) experiments of
K-dosed BP clearly showed that the band gap of BP can be completely closed,
and anisotropic Dirac fermions emerge as a result \cite{Kim:2015}.
Further studies have shown that
the electronic structure of BP is characterized in terms of
the anisotropic pseudospin structure
\cite{Choi:2015,Choi:2017}.

Pseudospin is an orbital degree of freedom that is usually used to describe
low-energy electronic structure of two-band systems.
The concept of pseudospin is particularly useful when crystal symmetry
endows certain structure in pseudospin texture.
Graphene, for example, has massless Dirac electrons with chiral pseudospin
structure owing to the symmetry of its honeycomb lattice \cite{Geim:2009}.
Then, this unique pseudospin structure gives rise to exotic phenomena called
Klein tunneling, where incident electrons tunnel through
potential barriers unimpeded \cite{Geim:2006,MacDonald:2012}.
Black phosphorus also has nontrivial pseudospin structure already in its
semiconducting phase, which comes from the glide symmetry \cite{Choi:2017}. 
Furthermore,
Dirac cones with chiral pseudospins emerge when the band gap of BP is inverted
\cite{Choi:2015,Choi:2017}.
Recently, the anisotropic pseudospin structure of BP has been experimentally
demonstrated using polarization-dependent ARPES measurements \cite{Kim:2020}.
However, unlike graphene, pseudospin-related transport phenomena in black
phosphorus are not yet fully explored \cite{Montambaux:2012,Louie:2017,Ang:2017,Knezevic:2017,stegmann:2019}.

\begin{figure}
  \centering
  \includegraphics[width=8.6cm]{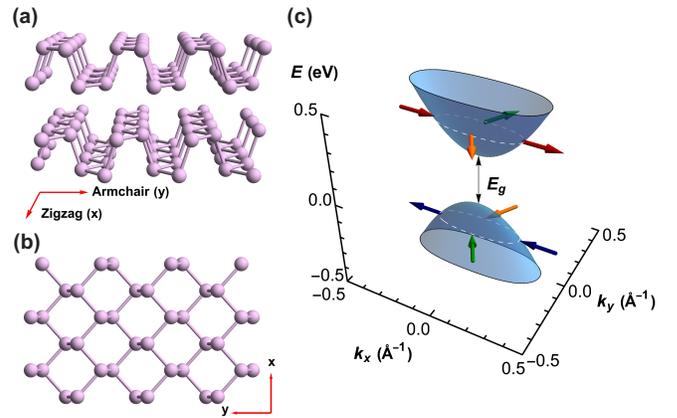}
  \caption {
    (a) Crystal structure of few-layer BP. The in-plane crystal
    structure is anisotropic with the zigzag ($x$) and armchair ($y$) directions.
    (b) Top view of the crystal structure.
    (c) Band structure and pseudospin texture of semiconducting BP ($E_g>0$),
    where $E_g$ indicates the band gap.
    Green, orange, red, and blue arrows represent pseudospin vectors.
  }
  \label{fig:crystal}
\end{figure}

\begin{figure*}
  \centering
  \includegraphics[width=0.95\textwidth]{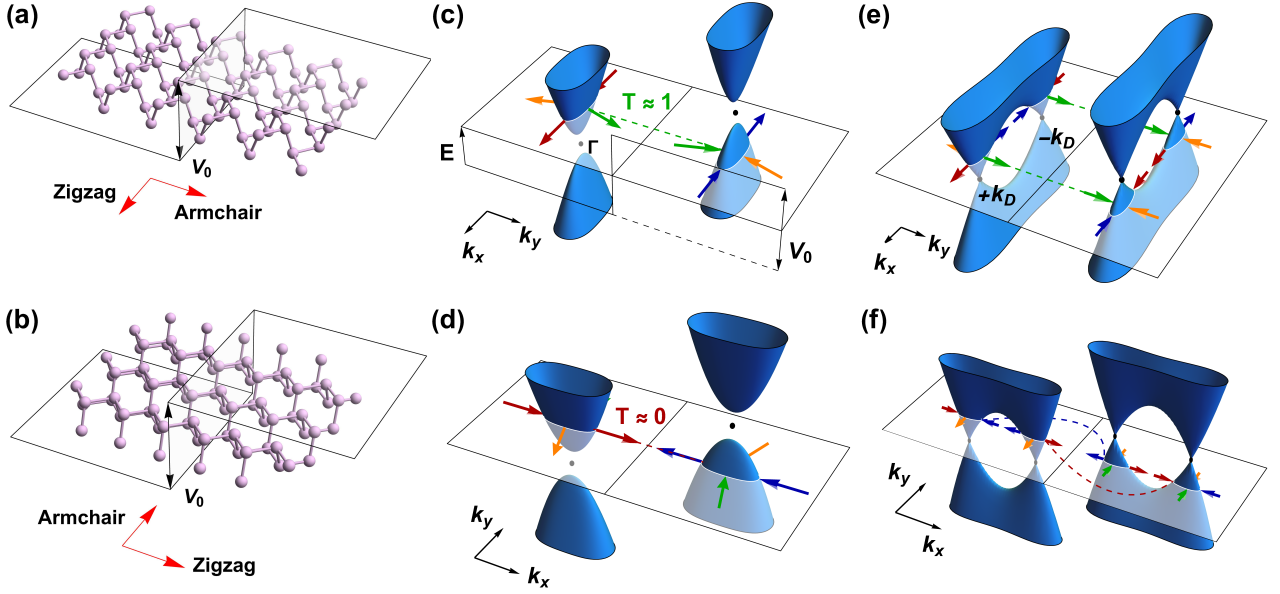}
  \caption {
    Black phosphorus junctions along (a) the armchair and
    (b) the zigzag direction.
    $V_0$ is the height of the potential energy step.
    Band alignment and pseudospin structure
    in (c) the armchair and (d) the zigzag junction
    with a normal band gap ($E_g>0$).
    $E$ is the energy of incident electrons and $V_0$ is larger than
    the band gap to allow overlap of bands in energy across the junction.
    In (c), pseudospins of incident electrons
    are nearly parallel to those of transmitted electrons so that
    transmission is almost unity.
    In (d), the interband tunneling is suppressed
    because pseudospins of incident and transmitted carriers
    are anti-parallel.
    (e),(f) Band alignment and pseudospin structure
    in inverted-gap ($E_g<0$) junctions for the armchair and zigzag direction,
    respectively.
    Dashed lines connect incident and transmitted states
    for the intravalley tunneling.
  }
  \label{fig:junctions}
\end{figure*}

In this work, we examine the pseudospin structure of BP and investigate
its effects on the interband tunneling properties across lateral BP junctions.
First, we describe the anisotropic pseudospin structure of BP using effective
two-band Hamiltonian, and emphasize important characteristics for
understanding interband tunneling properties.
Then, we explore the interband tunneling problems in lateral junctions formed
by a step potential orthogonal to each crystal axis.
The step potential can be controlled externally and it is related to,
for instance, a gate voltage in TFET.
Our calculations show that the interband tunneling is favorable in the armchair
direction while suppressed in the zigzag direction because of the pseudospin
matching conditions.
Lastly, we deal with the inverted-gap BP junctions and show that intravalley
tunneling is similar to the Klein tunneling for both junctions.

\section{Results and Discussion}
\subsection{Anisotropic Pseudospin Structure of few-layer BP}

To describe the electronic structure of few-layer BP near the $\Gamma$ points,
we use the following effective Hamiltonian in the pseudospin basis
\cite{Choi:2015,Choi:2017}.
\begin{equation}
  H(\mathbf{k}) =
    \left( \frac{E_g}{2} + \frac{\hbar^2 k_{x}^2}{2m^{*}} \right) \sigma_{x}
    + \hbar v_y k_{y} \sigma_{y},
\end{equation}
where the x and y directions are the zigzag and armchair directions,
respectively, $E_g$ is the band gap, which varies from 0.3 eV in bulk to 1.8 eV
in monolayer, $m^{*}= 1.42 m_e$ is the effective mass along the zigzag
direction ($m_e$ is the bare mass of electron in vacuum),
$v_y = 5.6 \times 10^5$ m/s is the velocity along the armchair
direction.
Here $\sigma_{x}\!=\!\big(\begin{array}{cc} 0 & 1 \\ 1 & 0 \end{array}\big)$ and
$\sigma_{y}\!=\!\big(\begin{array}{cc} 0& -i \\ i&0 \end{array}\big)$ are
the Pauli matrices that reflect the pseudospin degree of freedom.
This Hamiltonian can be derived straightforwardly 
by expressing the microscopic tight-binding 
Hamiltonian for black phosphorus as a polynomial up to the leading orders in
$k_x$ and $k_y$, which are the second order 
in $k_x$ and the first order in $k_y$ \cite{Choi:2017}.
This Hamiltonian is valid not only for positive $E_g$ but also for zero and
negative $E_g$.
While the form of the effective Hamiltonian is similar to that of deformed
honeycomb lattices \cite{Barnea:2010},
BP has opposite signs for the nearest-neighbor
hopping amplitudes along the zigzag and armchair directions.
In addition, the band gap of BP can be directly controlled by
the number of layers and vertical electric fields without changing
the anisotropy of hopping amplitudes \cite{Choi:2017}.

The anisotropic band dispersions in BP are given by
\begin{equation}
  \varepsilon_{\lambda} (\mathbf{k}) =
  \lambda \sqrt{
    \left(\frac{E_g}{2} + \frac{\hbar^2 k_{x}^2}{2m^{*}}\right)^2
    +\left(\hbar v_y k_y \right)^2
  }
  ,
\end{equation}
where $\lambda=\pm$ for the conduction and valence band, respectively.
Band dispersions are quadratic along the zigzag direction but hyperbolic along
the armchair direction.
The eigenstates are described by pseudospinors
\begin{equation}
  \chi_{\lambda} (\mathbf{k}) =
  \frac{1}{\sqrt{2}}\left(
  \begin{array}{cc}
    1 \\
    \lambda e^{i \theta_\mathbf{k}}
  \end{array}\right)
  ,
\end{equation}
where $\theta_\mathbf{k}=
\arg\left[
 (E_g/2 + \hbar^2 k_{x}^2/(2m^{*}))
 +i \hbar v_y k_y
\right]$ is the pseudospin angle.

Figure~\ref{fig:crystal}(c) shows the anisotropic pseudospin structure of BP.
The angle $\theta_\mathbf{k}$ is even in $k_x$ so two states in the same band
but with the opposite $k_x$ have the same pseudospin.
In contrast,
the armchair component of the pseudospin vector changes its sign when $k_y$ is
flipped because $\theta_\mathbf{k}$ is odd in $k_y$.
As a result, along the armchair direction,
the pseudospin of a conduction-band electron with positive $k_y$
makes a small angle with
the pseudospin of a valence-band electron with negative $k_y$ and vice versa.
In contrast, along the zigzag direction,
conduction and valence band remain nearly orthogonal
irrespective of $k_x$ and perfectly when $k_y=0$.
These properties are crucial
in understanding interband tunneling behaviors in lateral BP junctions.

\subsection{Pseudospin Structure of Lateral BP Junctions}
Now, we consider armchair and zigzag BP junctions as shown in
Figs.~\ref{fig:junctions}(a) and \ref{fig:junctions}(b), respectively.
The armchair (zigzag) junction is formed by a step potential of height $V_0$
such that the junction interface is normal to the armchair (zigzag) direction.
$V_0$ should be larger than the band gap
to make the conduction band in one side and
the valence band in the other side be overlapped in energy.

Before quantitative calculations,
we can qualitatively understand interband tunneling properties
by looking at the alignment of band and pseudospin.
Figure~\ref{fig:junctions}(c) shows the interband
tunneling mechanism for the armchair junction with a normal band gap ($E_g>0$).
Pseudospins of incident electrons are nearly parallel
with those of the transmitted electrons.
In contrast, reflected electrons have nearly antiparallel pseudospins.
As a result, we can expect the interband transmission $T$ of nearly unity.
The reason why $T$ is not exactly 1 is because pseudospins are not perfectly
aligned. Even when $k_{x}=0$, the zigzag component of pseudospin is still
nonzero due to the band gap. In fact, $T$ becomes exactly 1
when the band gap is closed.
Tunneling in the armchair direction of BP strongly resembles the Klein tunneling
in monolayer graphene \cite{Geim:2006}.

\begin{figure}
    \centering
    \includegraphics[width=8.6cm]{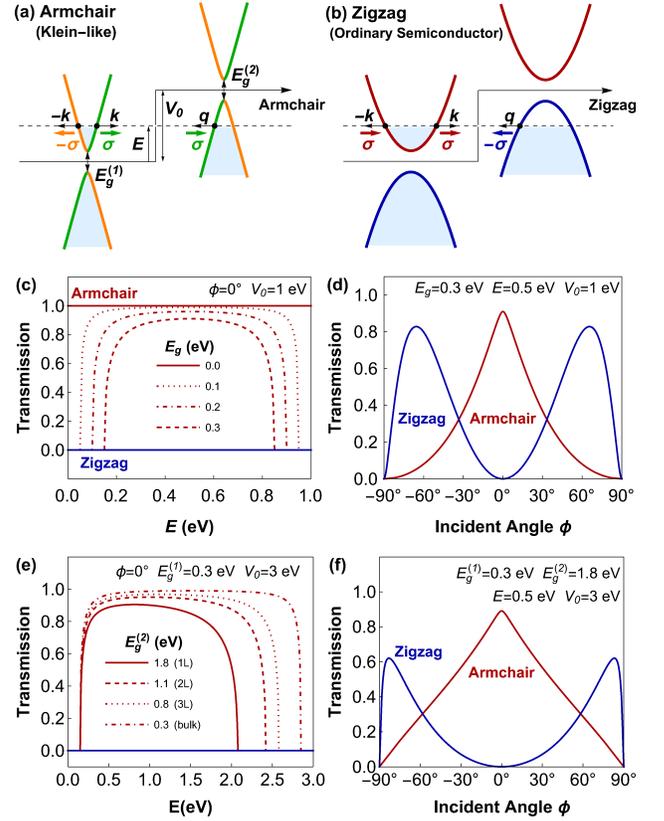}
    \caption {
      (a) Schematic view of the normal incidence in the armchair junction,
      where the left (right) region has the band gap $E_g^{(1)}$ ($E_g^{(2)}$).
      Incident electrons in the conduction band
      with the momentum $\mathbf{k}$ are transmitted to
      electrons in the valence band with the momentum $\mathbf{q}$.
      Klein-like tunneling
      occurs due to nearly aligned pseudospins between
      incident and transmitted electrons.
      (b) Normal incidence in the zigzag junction.
      Total reflection occurs because of exactly antiparallel pseudospins
      between incident and transmitted electrons.
      (c) Transmission probability in homogeneous junctions
      ($E_g^{(1)}=E_g^{(2)} = E_g$) as a function of the incident energy $E$.
      Red lines are for armchair junctions,
      where four different band gaps are considered.
      The blue line is for the zigzag junction,
      where the normal transmission is always zero
      irrespective of the band gap size.
      (d) Incident-angle dependence of the transmission probability
      for homogeneous armchair (red) and zigzag (blue) junctions.
      (e) Normal incidence and (f) incident-angle dependence
      in heterojunctions ($E_g^{(1)}\neq E_g^{(2)}$).
      In (d) and (f), the incident angle $\phi$ represents the direction of
      the group velocity of the incident carrier
      with respect to the normal direction of the interface, 
      with $\phi = 0^\circ$ for the normal incidence.
    }
    \label{fig:tunnel-normal}
\end{figure}

Figure~\ref{fig:junctions}(d) shows a contrasting situation in the zigzag junction.
Pseudospins of incident electrons are antiparallel to transmitted
electrons, but parallel to reflected electrons in the same side.
Thus, the interband tunneling is suppressed, and exactly zero
when $k_{y}=0$. Tunneling in the zigzag direction of BP is similar to
ordinary semiconductor junctions, where interband tunneling is usually
not efficient because conduction and valence bands are orthogonal.
In short, interband tunneling in BP is favorable in the armchair junction
but suppressed in the zigzag junction due to the pseudospin matching
condition.

Owing to its unique pseudospin structure, black phosphorus undergoes
topological phase transition from a normal semiconductor to 2D Dirac semimetal
when the band gap is inverted \cite{Choi:2015,Choi:2017,Kim:2017}.
The low-energy electronic structure is then characterized by two Dirac cones.
Figures~\ref{fig:junctions}(e) and (f) show the band alignment and pseudospin
structure of inverted-gap BP junctions.
The pseudospin structure in both armchair and zigzag junctions favors
transmission within the same valley.
In fact, as we will show below,
intravalley tunneling in the armchair junction
is the perfect Klein tunneling.
On the contrary, intervalley tunneling is strictly forbidden in the armchair junction
because of the momentum conservation,
and is very weak also in the zigzag junction
because of the pseudospin mismatch.

\subsection{Interband Tunneling in Semiconducting BP Junctions}

For quantitative analysis,
we calculate transmission and reflection probabilities
by matching scattering-state wavefunctions
at the interface of the junction.
First, we consider interband tunneling problem in the armchair junction
for a given incident energy $E$: $H\psi=E\psi$.
Since there are two propagating modes in each side of the junction,
we can write down the scattering-state wavefunction as
\begin{equation}
  \psi (x,y) \! = \!
  \left\{ \!
    \begin{array}{ll} \!
      \begin{array}{l}
      e^{i k_x x } e^{i k_y y} \chi_{+} (k_x, k_y)\\
        + r \;\!  e^{i k_x x } e^{- i k_y y} \chi_{+} (k_x, -k_y),
      \end{array}
      &
      \!\textrm{for}\;y < 0
      \\[0.5cm]
      t \; \! e^{i q_x x } e^{i q_y y}
      \chi_{-} (q_x, q_y),
      &
      \!\textrm{for}\;y>0
      ,
    \end{array}
  \right.
\end{equation}
where $r$ and $t$ are reflection and transmission amplitudes, respectively.
The incident electron has the momentum
$\hbar k_x$ and
$\hbar k_y=\frac{1}{v_y}
\sqrt{E^2-(\frac{E_g^{(1)}}{2}+\frac{\hbar^2 k_x^2}{2m^{*}})^2}$,
the transmitted electron has $\hbar q_x=\hbar k_x$ and
$\hbar q_y=-\frac{1}{v_y}
\sqrt{(E\!-\!V_0)^2-(\frac{E_g^{(2)}}{2}+\frac{\hbar^2 k_x^2}{2m^{*}})^2}$,
where $E_g^{(1)}$ and $E_g^{(2)}$ are band gaps in two sides of the junction,
which can be different, in general.
Then, the transmission probability $T$ is given by
\begin{equation}
T= 1-|r|^2=
-\frac{ 2 \sin(\theta_\mathbf{k}) \sin(\theta_\mathbf{q}) }
{ 1+\cos(\theta_\mathbf{k}+\theta_\mathbf{q}) },
\end{equation}
where $\theta_\mathbf{k}$ and $\theta_\mathbf{q}$ are pseudospin angles defined
above.
In the case of a normal incidence ($k_x=0$) with zero band gap ($E^{(1)}_g=E^{(2)}_g=0$),
$\theta_\mathbf{k}=\pi/2$ and $\theta_\mathbf{q}=-\pi/2$, resulting in $T=1$.
The perfect transmission of normally incident electrons, independent of the
step height, is the signature of the Klein tunneling,
which happens in monolayer graphene \cite{Geim:2006}.
With $E_g^{(1)}=E_g^{(2)}>0$,
the pseudospins are tilted in the zigzag direction so that the
maximum transmission becomes less than 1 even for the normal 
incidence, as shown in Figs.~\ref{fig:tunnel-normal}(c) and \ref{fig:tunnel-normal}(d).
In this case, the maximum transmission occurs when the incident energy is 
maximally away from both conduction- and valence-band edges 
[Fig.~\ref{fig:tunnel-normal}(c)].

Next, we consider the zigzag junction, where the step potential is at $x=0$.
In this case, there are four eigenstates of the Hamiltonian for a given
energy $E$. Two of them are propagating modes and the other two are evanescent modes
which are localized at and propagate along the junction interface.
Thus, the scattering-state wavefunction is
\begin{equation}
  \psi (x,y) \! = \!
  \left\{ \!
    \begin{array}{ll} \!
      \begin{array}{l}
        e^{i k_x x } e^{i k_y y} \chi_{+} (k_x, k_y) \\
        + r \; e^{-i k_x x } e^{i k_y y} \chi_{+} (-k_x, k_y) \\
        + a \; e^{ K_x x } e^{ i k_y y} \chi_{+} (-i K_x, k_y),
      \end{array}
      &
      \! \textrm{for}\;x<0 \\[0.5cm]
   \!   \begin{array}{l}
       t\; e^{i q_x x} e^{i q_y y} \chi_{-} (q_x, q_y)\\
       + b \; e^{ - Q_x x } e^{ i q_y y} \chi_{-} (i Q_x, q_y),
      \end{array}
      &
   \!   \textrm{for}\;x>0
    \end{array}
  \right.
\end{equation}
where $k_y=q_y$ and decaying parameters for evanescent waves are
$\hbar K_x=\sqrt{2m^{*}E_g^{(1)} + (\hbar k_x)^2}$
and $\hbar Q_x=\sqrt{2m^{*}E_g^{(2)} + (\hbar q_x)^2}$.

\begin{figure}
  \centering
 \includegraphics[width=8.6cm]{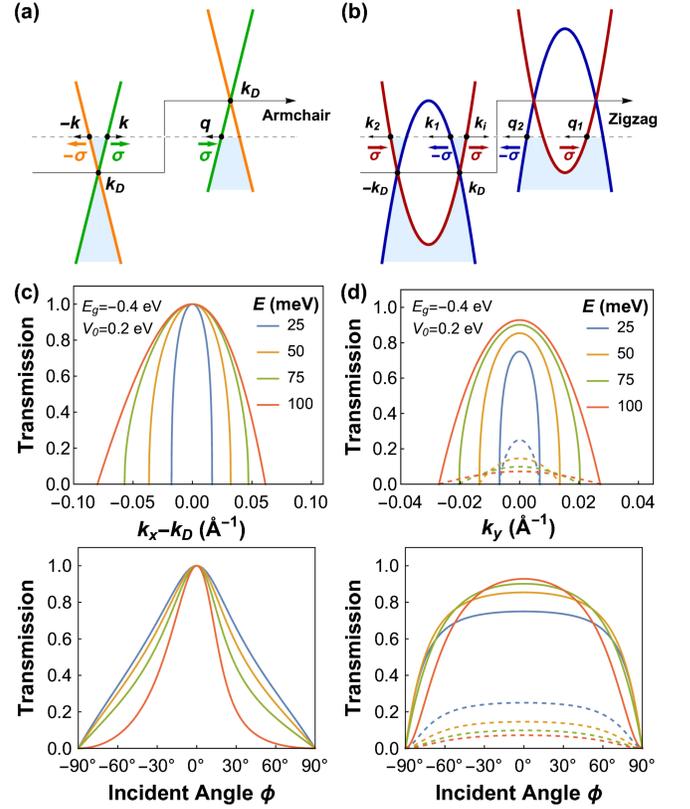}
  \caption {
    Schematic view of the normal incidence in (a) the armchair junction
    for the valley at $+k_D$ and (b) the zigzag junction
    with an inverted band gap ($E_g<0$).
    (c) Intravalley transmission in the armchair junction plotted as 
    a function of the zigzag momentum (the upper panel) and 
    the incident angle (the lower panel). 
    Perfect transmission occurs at $k_x=\pm k_D$.
    (d) Intravalley transmission in the zigzag junction plotted as a function of
    the armchair momentum (the upper pannel) and the incident angle (the lower pannel).
    In (c) and (d), blue, orange, green, and red solid lines represent 
transmission
    for incident energies of 25, 50, 75, and 100 meV, respectively.
    In (d), dashed lines indicate reflection probability to the other valley in the
    same side of the junction for incident energies of solid lines.
  }
  \label{fig:tunnel-inverted}
\end{figure}

Figure~\ref{fig:tunnel-normal}(c) shows that
the interband transmission for the normal incidence in the zigzag junction
is exactly 0 irrespective of the incident energy.
This is because pseudospins of incident and transmitted carriers are
exactly antiparallel.
For general incident angles, however,
there are certain windows where the interband transmission becomes sizable
[Fig.~\ref{fig:tunnel-normal}(d)].
This nontrivial angular dependence of the transmission is captured only when
evanescent waves are taken into account.

Since the band gap of BP sensitively depends on
the number of layers or external electric fields
\cite{Choi:2015,Zunger:2015,Fengnian:2017},
we also consider
the case that two sides of the junction have different band gaps
($E_g^{(1)}\neq E_g^{(2)}$).
Figures~\ref{fig:tunnel-normal}(e) and \ref{fig:tunnel-normal}(f)
show transmission properties in heterojunctions between
bulk BP ($E_g^{(1)}$ = 0.3 eV) and mono-, bi-, and, tri-layer BP
($E_g^{(2)}$ = 1.8, 1.1, and, 0.8 eV, respectively).
Band-gap values are taken from Ref.~\citenum{Wang:2017}.
The obtained transmission is substantially large in all considered bulk-layer
armchair junctions whenever the incident energy is not too close to the band edges,
while zigzag junctions allow no normal transmission.
The incident-angle dependence is also similar to the homogeneous junction,
except that the maximum transmission in the zigzag junction becomes much smaller
whereas that in the armchair junction is relatively unaffected.

\subsection{Interband Tunneling in Inverted-Gap BP Junctions}

Lastly, we calculate transmission properties of inverted-gap BP junctions,
where both sides of junctions have the same negative band gap.
When the band gap is inverted in BP, two Dirac cones are separated in the momentum space
along the zigzag direction. Thus, in the armchair junction,
electrons in one valley cannot transmit to the other valley
because of the momentum conservation.
Within the same valley, the pseudospin structure enforces that
the perfect tunneling occurs at the normal incidence,
which is the Klein tunneling
[Fig.~\ref{fig:tunnel-inverted}(c)].
In the zigzag junction, unless the incidence is normal to the interface, 
intervalley transmission is not strictly forbidden
but it is two orders of magnitude smaller than intravalley transmission
since the pseudospin between different valleys are almost opposite.
Meanwhile, intravalley transmission is not perfect, even for the normal incidence, because
the pseudospin structure allows reflection to the other valley in the same
side of the junction [Fig.~\ref{fig:tunnel-inverted}(d)].
Here the intervalley reflection is substantial because the
potential change is abrupt at the interface. If the potential changes 
from 0 to $V_0$ gradually in a region much wider than the unitcell length,
the intervalley reflection becomes negligible
and the transmission at the normal incidence becomes almost perfect.

\section{Conclusion}
To summarize, we have investigated the pseudospin structure and its role for
the interband tunneling in lateral black phosphorus junctions.
In semiconducting BP, the pseudospin structure is highly anisotropic
so that the pseudospin angle is an even function with respect to the zigzag
momentum while odd in the armchair momentum.
In other words,
orbital characters of the conduction and valence bands do not change
when the zigzag momentum is flipped, but the conduction band states
with the positive armchair momentum have strong overlap with the valence band
states with the negative armchair momentum, and vice versa.
Therefore, when a lateral junction is formed in the armchair direction,
pseudospins of incident and transmitted electrons are nearly
aligned. So the interband tunneling is highly favorable in the armchair
junction. In contrary, the interband tunneling in the zigzag junction is
suppressed due to misaligned pseudospins. It is zero for the normal
incidence while there are certain ranges of incident angles with
sizable transmission.
We also showed that pseudospin structure in the inverted-gap BP junctions
facilitate the intravalley tunneling in both armchair and zigzag junctions.
Our results provide insights into the mechanism of tunneling-based
transport with pseudospins in tunable gapped semi-Dirac materials.

\begin{acknowledgements}
  This work was supported by NRF of Korea (Grant No. 2020R1A2C3013673)
  and KISTI supercomputing center (Project No. KSC-2019-CRE-0195).
  Y.W.C. acknowledges support from NRF of Korea
  (Global Ph.D. Fellowship Program NRF-2017H1A2A1042152).
\end{acknowledgements}

\providecommand{\newblock}{}

\end{document}